\documentclass[twocolumn,showpacs,preprintnumbers,amsmath,amssymb,prl]{revtex4}

\usepackage{graphicx}
\usepackage{bm}

\begin{document}

\title{Field-Induced Thermal Metal-to-Insulator Transition in Underdoped LSCO}

\author{D.G.~Hawthorn$^1$, R.W.~Hill$^1$, C.~Proust$^{1,*}$, F.~Ronning$^1$, Mike~Sutherland$^1$, Etienne~Boaknin$^1$,
C.~Lupien$^{1,\dagger }$, M.A.~Tanatar$^1$,
Johnpierre~Paglione$^1$, S.~Wakimoto$^1$, H.~Zhang$^1$, Louis~Taillefer$^{1,\ddagger }$, T.~Kimura$^2$, M.~Nohara$^2$,
H.~Takagi$^2$ and N.E.~Hussey$^3$}

\affiliation{$^1$Department of Physics, University of
Toronto, Toronto, Ontario, Canada}

\affiliation{$^2$Department of Advanced Materials Science, Graduate School of Frontier Sciences, University of
Tokyo, Hongo 7-3-1, Bunkyo-ku, Tokyo 113-8656, Japan}

\affiliation{$^3$H.H. Wills Physics Laboratory, University of Bristol, Bristol, United Kingdom}

\date{\today}

\begin{abstract}
The transport of heat and charge in cuprates was measured in single crystals of La$_{2-x}$Sr$_x$CuO$_{4+\delta }$ (LSCO)
across the doping phase diagram
at low temperatures. In underdoped LSCO, the thermal conductivity is found to decrease with increasing magnetic
field in the $T \rightarrow 0$ limit,
in striking contrast to the increase observed in all superconductors, including cuprates at higher doping.
In heavily-underdoped LSCO, where superconductivity can be entirely suppressed with an applied magnetic field,
we show that a novel thermal metal-to-insulator transition takes place upon going from the superconducting state
to the field-induced normal state.

\end{abstract}


\maketitle

In underdoped La$_{2-x}$Sr$_x$CuO$_{4+\delta}$ (LSCO), resistivity measurements have revealed the field-induced
normal state to be a charge insulator \cite{Ando}.
On the other hand, the superconducting state of underdoped LSCO is a thermal metal,
in the sense that there is a clear $T$-linear contribution to the thermal conductivity at $T \rightarrow 0$
\cite{Sutherland,Takeya}.
Given that in all superconductors investigated to date (including cuprates) heat transport at $T \rightarrow 0$
is always seen to increase as one goes from the superconducting state to the field-induced normal state,
these two observations point to a violation of the Wiedemann-Franz law in underdoped cuprates. Note that this universal law
is violated in the electron-doped cuprate Pr$_{2-x}$Ce$_x$CuO$_{4+\delta}$ (PCCO) at optimal
doping \cite{Hill}, in
that low-temperature heat conduction was found to exceed the expected charge conduction by a factor of approximately two.
However, the law is recovered in the strongly overdoped regime \cite{Proust}.

In this Letter, we show the natural assumption that heat conduction will increase upon going
from the superconducting state to the field-induced normal state to be incorrect in underdoped LSCO.
Indeed, in the $T \rightarrow 0$
limit the thermal conductivity {\it decreases} in the vortex state and the residual linear term drops to
a value below our resolution limit in the field-induced normal state.
This argues strongly for a thermally insulating normal state and reveals a novel thermal metal-to-insulator
transition.
These findings shed new light on the nature of the intriguing state of underdoped cuprates above $H_{c2}$, for which several
proposals have been put forward recently, including stripe order \cite{Carlson2}, {\it d}-density wave order \cite{Chakravarty} and
a Wigner crystal of {\it d}-wave hole pairs \cite{ChenHD}.
As for the Wiedemann-Franz law, a strict test will prove elusive as
transport in both charge and heat channels show insulating behavior (in
the normal state).

\begin{figure*}[tbh]
\centering
\resizebox{6.3in}{!}{\includegraphics[angle=-90]{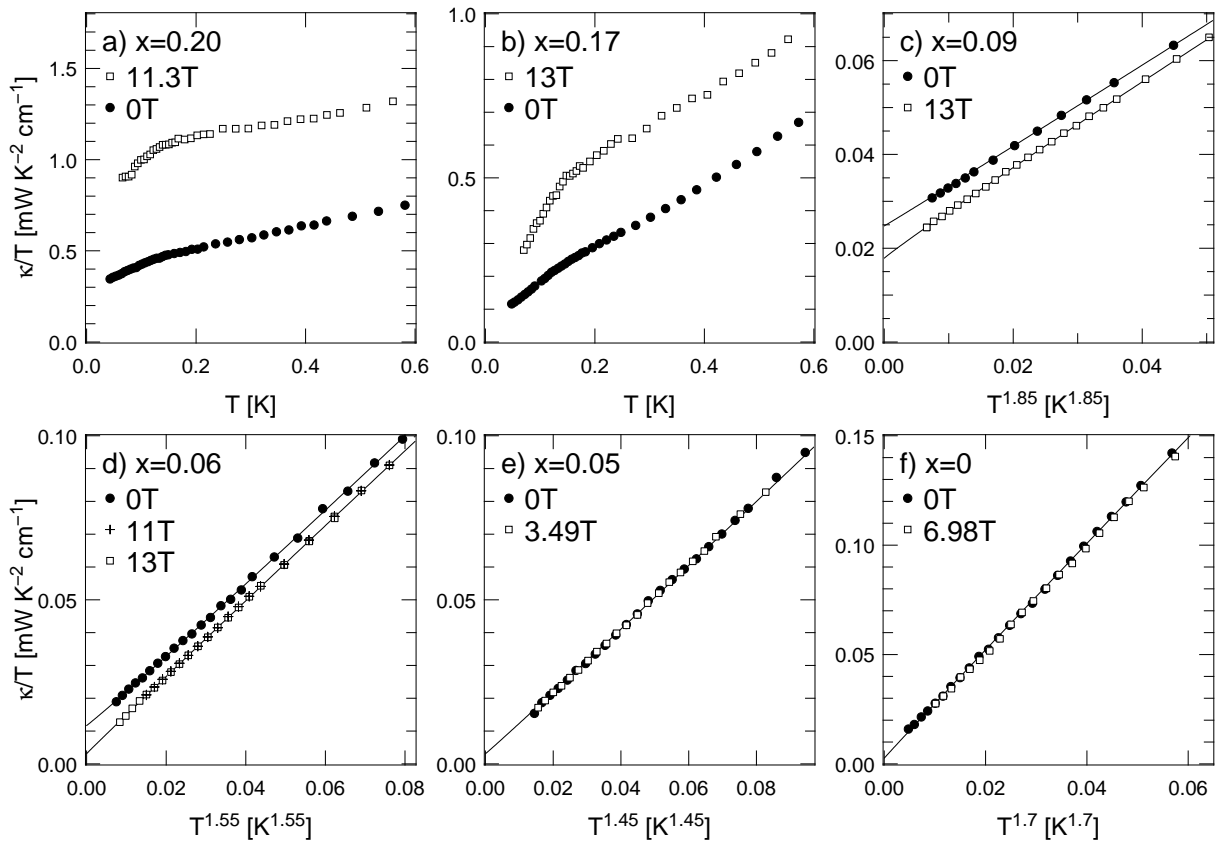}}
\caption{\label{fig:kappa}
$\kappa/T $ vs. $T^{\alpha -1}$ for La$_{2-x}$Sr$_x$CuO$_{4+\delta}$ with $x$ as shown.  The lines are fits to Eq.~\ref{eq:kappafit}.}
\end{figure*}
Measurements of the thermal conductivity ($\kappa $) were performed down to 40 mK in magnetic fields up to 13 T
on single crystals of  La$_{2-x}$Sr$_x$CuO$_{4+\delta}$ with Sr doping $x=0$, 0.05, 0.06, 0.09, 0.17 and 0.20.
All samples were grown in an image furnace using the travelling solvent floating zone technique.
Sample
$x=0$ was annealed at
700~$^o$C in flowing argon gas overnight, in order to set the oxygen content at the stoichiometric O$_4$.
Other samples were not annealed after growth.
The samples were
oriented using Laue diffraction and cut into rectangular samples of typical
 dimensions of length equal to 1-2 mm, thickness equal to 0.1-0.3 mm and width equal to 0.4-1 mm, with
the current flowing along the $a$ axis of the tetragonal unit cell.
Electrical contacts to the samples were made
using silver epoxy cured at 500 $^o$C in flowing O$_2$.
(In the undoped sample, argon was used instead of oxygen.)
Silver wires were then attached to the contact pads using silver epoxy.  The error in
the geometric factor resulting from the finite width of the contact pads is less than 15\%.
The resistively determined transition temperature $T_c$ is
5.5,
16, 34 and 33.5~K for the $x=0.06$
0.09, 0.17 and 0.20 samples, respectively,
where $T_c$ is defined as the temperature where $\rho =0$.
(Samples with $x=0$ and 0.05 are not superconducting.)
Thermal conductivity measurements on the same samples in zero field are discussed in Ref.~\cite{Sutherland}, along with
details of the measurement technique and data analysis.
Resistivity was also measured on these samples using the same contacts.

In Fig.~\ref{fig:kappa} the thermal conductivity is plotted as $\kappa/T$ vs. $T^{\alpha -1}$,
where $\alpha $ is a free fitting parameter.
This type of plot is used to separate the electronic
($\kappa_{el}$)
and lattice
($\kappa_{ph}$)
contributions to $\kappa $ by making use of their different
power-law temperature dependences in the $T \rightarrow 0$ limit.
In the limit $T \to 0$ $\kappa_{el}$ is linear in $T$ for a $d$-wave superconductor on
account of nodal quasiparticle excitations \cite{Durst,Graf}.
Quite generally, a linear contribution to $\kappa $ at $T\to 0$
is direct evidence for fermionic excitations.
The phonon contribution can be modeled as
$\kappa_{ph} \propto T^\alpha $ for phonons limited to scattering from the boundaries of the sample.
Note that this in general differs from the simple $T^3$
dependence assumed in previous work, a departure which arises from
specular reflection of phonons off smooth crystal faces.
Thus, $\kappa_{el}$ and $\kappa_{ph}$ can be separated by fitting the
data at low-temperatures to

\begin{equation}
\frac{\kappa }{T} = \frac{\kappa_o}{T} + BT^{\alpha -1} .
\label{eq:kappafit}
\end{equation}

The two distinct contributions are identified in Fig.~\ref{fig:kappa} as the intercept and slope of the curves, respectively, when plotting the data as
$\kappa /T$ vs. $T^{\alpha -1}$.
Eq.~\ref{eq:kappafit} provides an excellent fit to the data for the underdoped samples.
Quite generally for almost any material and sample we have investigated,
this fitting procedure is a definite improvement upon the standard approach
of fitting to $\kappa /T = \kappa_o/T + BT^2$, which requires one to typically only make use of data
below 150 mK and often results in an overestimate of $\kappa_o/T$ (see Ref.~\cite{Sutherland}).

Although Eq.~\ref{eq:kappafit} works well for the underdoped samples,
it does not provide a good description of the $x=0.20$ and 0.17 data.  For
this
 reason, the data for these two more
highly doped samples is
plotted simply vs. $T$ in Fig.~\ref{fig:kappa}.  The inapplicability of
the
fit in these two cases
is due to a downturn in the data below about 150 mK, as observable in
Figs.~\ref{fig:kappa}a and \ref{fig:kappa}b.
These downturns, however, only impact upon the $T$ dependence and not the $H$
dependence of $\kappa $.
Thus, although the origin of the downturns remains uncertain
the field dependence can be understood independently of the $T$ dependence,
which is left as a topic of future investigation.

\begin{figure}[htb]
\centering
\resizebox{\columnwidth }{!}{\includegraphics{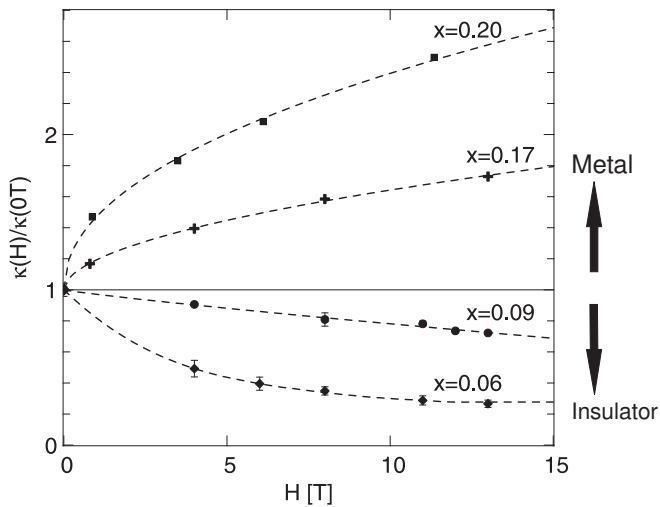}}
\caption{\label{fig:kvsH}$\kappa_0(H)/T$ (normalized to $\kappa(0\mathrm{T})/T$) vs $H$.
For $x=0.20$ and 0.17, an isotherm
of $\kappa(H)/T $ at 60 mK is plotted instead of $\kappa_o/T$.  The dashed lines are guides to the eye.  }
\end{figure}

Having described our analysis of the data, several observations can be made.
Firstly, in zero field (solid circles) the data reproduces the results of Refs.~\cite{Takeya} and \cite{Sutherland}
whereby a finite residual linear term in $\kappa(T)$ is resolved for $x \geq 0.06$.
This proves the existence of delocalized zero-energy quasiparticles throughout the superconducting region.
(Among other implications, this essentially rules out the possibility of a $d+ix$ order parameter, where $x=s$ or $d$; see
\cite{Sutherland,Proust}.)
In other words, the $d$-wave superconducting state is a thermal metal at all dopings (see also \cite{Sutherland}).
Outside the superconducting region,
{\it i.e.} below $x=0.055$,
the residual linear term becomes extremely small. The power-law fitting procedure described above
to extrapolate
to $T=0$ yields a value of $\kappa_0/T = 3~\mu$W~K$^{-2}$~cm$^{-1}$ for both the $x=0$ and $x=0.05$ samples.
Now, even though the power-law fit does provide a good description of the data (see Figs.~1e and 1f),
all the way up to 0.4~K,
the fact that $\kappa_0/T$ is 5 times smaller than the value of $\kappa/T$ at the lowest data point (40~mK)
means that one has to view the extrapolated value with caution.
The conservative position is to assume that the parent compound $x=0$ is a heat insulator
as well as a charge insulator, to regard this minute linear term of
$3~\mu$W~K$^{-2}$~cm$^{-1}$ as the resolution limit
of our technique
(for this series of samples) \cite{endnote1}, and to treat the $x=0$ data as our reference (for an insulating state in LSCO crystals).
We emphasize that the $x=0.05$ sample is no more
conductive than the parent compound (see Fig.~2),
and hence is also taken to be a thermal insulator. By contrast, the linear term in the
$x=0.06$ sample (at zero field), of magnitude $12~\mu$W~K$^{-2}$~cm$^{-1}$,
is clearly above the reference limit (by a factor 4)
and is thus unambiguously a thermal metal.

\begin{figure}
\centering
\vspace{8pt}
\resizebox{\columnwidth}{!}{\includegraphics{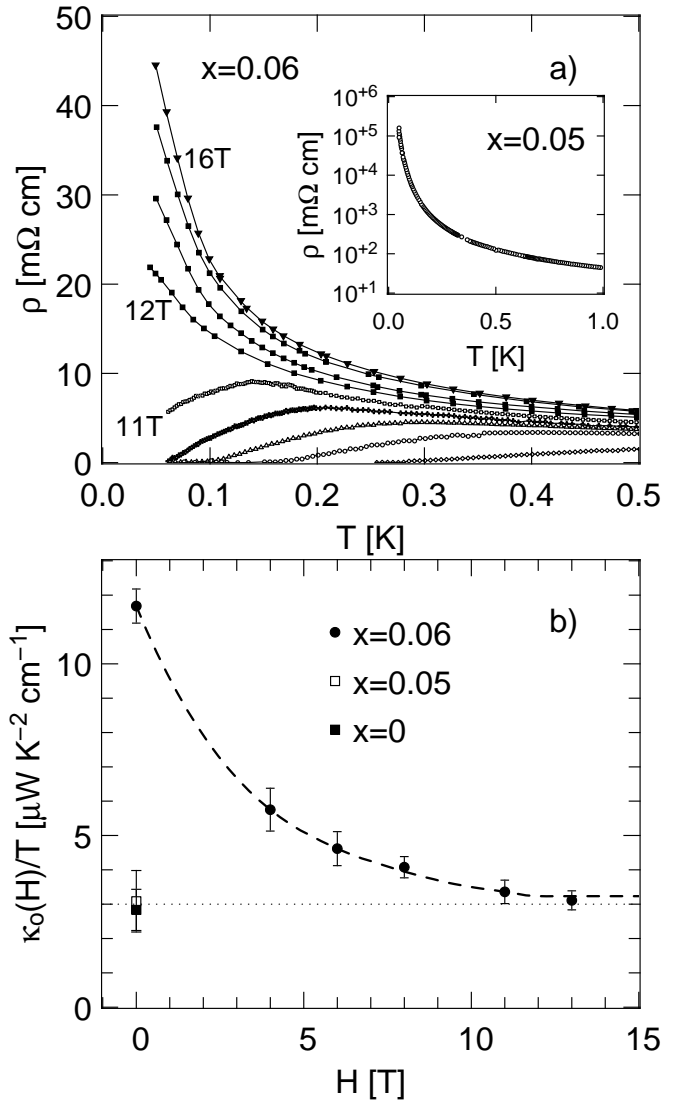}}
\caption{\label{fig:reswkvsH} a) Low-temperature resistivity of LSCO with $x=0.06$ in magnetic fields
of 6, 8, 9, 10, 11, 12, 13, 15 and 16~T.
By 12~T the superconducting
 transition has been suppressed to below 40 mK.  Inset:  Resistivity of LSCO with $x=0.05$ on a semi-log scale.
 b) $\kappa_o(H)/T$ vs. $H$ for LSCO with $x=0.06$.  $\kappa_o/T$ is also
 shown for $x = 0$ and 0.05 at zero field.  The dotted line represents the estimated resolution of our experiment
 (see text).
 The error bars are statistical errors in the fitted
 values of $\kappa_o/T$ and do not include errors in the geometric factors
(which do not change with field) or systematic errors in the fitting procedure (discussed in the text).
The dashed line is a guide to the eye.}
\end{figure}

This brings us to the second, and principal observation that $\kappa$ {\it decreases with increasing field for
the underdoped samples} ($x=0.06$ and 0.09).
In Fig.~\ref{fig:kvsH}, the field evolution of $\kappa_{el}/T$ is shown at fields intermediate between 0 and 13~T,
as $\kappa_o/T$ for LSCO $x=0.06$ and 0.09, and as isotherms of $\kappa/T$ at 60 mK for $x=0.17$ and 0.20
(because of the previously stated difficulties in extrapolating to $T$=0 the data for these two samples).
By contrast to the underdoped samples,
the electronic heat conductivity in the more highly doped samples ($x=0.17$ and 0.20) increases with field, as it does
in all known superconductors at $T \to 0$ \cite{endnote2}. This increase in $\kappa $ at 60 mK is qualitatively consistent with the $T=0$
field dependence observed in optimally doped YBCO \cite{Chiao2}, which is roughly $\sqrt{H}$ and described by semi-classical models \cite{Kubert,Vekhter}.
Note that $\kappa$ is totally independent of magnetic field in our reference sample ($x=0$),
as in the $x=0.05$ sample.
This shows that field dependence is a property of the superconducting state. We can therefore
use this
criterion to establish that the non-superconducting normal state is reached in the bulk
by 11 T in sample $x=0.06$. Indeed, as seen in Fig.~1d, a further increase of the field to 13 T
causes no further change in $\kappa$.
This claim is supported by resistivity measurements, shown in Fig.~\ref{fig:reswkvsH}a,
where the resistive onset of superconductivity is entirely absent for fields of 12~T and above (down to 40 mK).
We take this as an additional indication that the field-induced (non-superconducting) normal state has been reached
by 13 T at $x=0.06$ (in the bulk).
A zoom on the $x=0.06$ data is shown in Fig.~\ref{fig:reswkvsH}b, where we can see that $\kappa_0/T$ drops
by a factor 4 from $H=0$ to $H=13$~T, where it reaches a value equal to that of the reference sample, namely
$\kappa_0/T = 3~\mu$W~K$^{-2}$~cm$^{-1}$. We conclude that the field-induced normal state in underdoped LSCO is a
thermal insulator. {\em This implies the existence of an unprecedented kind of thermal metal-to-insulator transition.} The
superconducting state is a thermal metal by virtue of its delocalized nodal quasiparticles, while the field-induced
normal state in the same sample is a thermal insulator, with either no fermionic excitations or localized fermionic
excitations. The well-known crossover from charge metal to charge insulator, identified as the change from a positive
$d\rho/dT$ at low $T$ near and above optimal doping to a negative $d\rho/dT$ in underdoped LSCO \cite{Ando}, now finds
a parallel in the heat sector where the metal-insulator
crossover is identified as the change from a positive $d\kappa/dH$ at low $T$ to a negative $d\kappa/dH$
(see Fig.~\ref{fig:kvsH}).

The fundamental question is: what does this imply for the nature of the field-induced normal state in underdoped
cuprates? The answer to this question depends on the role of disorder. Is the ``normal'' state intrinsically metallic,
with fermionic excitations which would be delocalized in the absence of disorder?
If so, an explanation must be found for why quasiparticles escape localization in the superconducting state.
Alternatively, is the normal state intrinsically insulating,
with no fermionic excitations (at low energy)?

A tantalizing possibility is that the thermal metal-to-insulator transition is a signature of a competing order that co-exists with superconductivity
(see for example \cite{Chakravarty,Demler,Kivelson}).  In this picture the application of a field suppresses superconductivity and thereby
allows the competing phase to define the ground state excitations.
Some candidates for the competing phase are thermal insulators ({\it e.g.} antiferromagnetic order, Wigner crystal),
while others are thermal metals (staggered flux phase, spin-charge separation).
The doping evolution of the field dependence of $\kappa $ with increasing doping follows naturally from the
decreasing importance of the competing order.
For instance by $x=0.17$ the ground state may have already undergone a quantum critical transition to a state where the only ordered state is
that of the {\it d}-wave superconductor, thus accounting for the recovery of the conventional increase in $\kappa $ with field.
One way to possibly distinguish between the two scenarios for the origin of insulating behavior
is to reduce the level of disorder and see whether a thermal metal is uncovered.

In closing we mention that because of the insulating character of both heat and charge conduction in $x=0.06$ at 13~T, it
is technically not possible to perform a real test of the Wiedemann-Franz law. Consequently,
the possibility that the two conductivities cease to be equal as they are in overdoped Tl-2201 \cite{Proust} but rather
diverge as they do in PCCO \cite{Hill} is still an open question.

In summary, we have observed in underdoped LSCO a decrease in thermal conductivity with magnetic field
upon going from the superconducting state to the field-induced normal state.
 We show that this result is due to a novel
thermal metal-to-insulator transition.  By contrast, for optimally and overdoped LSCO the thermal conductivity
increases with field, as it does in all other known superconductors.
These results impose clear constraints on models of the field-induced normal state of underdoped cuprates, and are expected
to help distinguish between various proposed phases that are postulated to compete with superconductivity in this part of
the phase diagram.
To this end, it will be important to understand the role of disorder.

\begin{acknowledgments}
This work was supported by the Canadian Institute for Advanced Research and NSERC of Canada.
We would like to thank H. Dabkowska, B.D. Gaulin, G. Luke and K. Hirota for use of their image furnaces and P. Fournier and W.A. MacFarlane
for help with the measurements.
\end{acknowledgments}

$*$ Present address: Laboratoire National des Champs
Magn\'etiques Puls\'es, 31432 Toulouse, France.

$\dagger $ Present address: Department of Physics, Cornell University, Ithaca, NY 14853, USA.

$\ddagger $ Present address: D\'epartment de Physique, Universit\'e de Sherbrooke, Sherbrooke, Qu\'ebec, Canada.

\bibliography{fieldpaper_short}

\begin{thebibliography}{17}
\bibitem[{end({\natexlab{a}})}]{endnote1}
\bibinfo{note}{In fitting the data using Eq.~\ref{eq:kappafit}, an extraneous
  contribution to $\kappa_0/T$ will occur whenever the actual phonon
  conductivity deviates from our model of $BT^\alpha $. This is the likely
  origin of the minute linear terms in the $x=0$ and 0.05 samples. Note that an
  error of this sort provides an offset to $\kappa_0/T$ but does not impair our
  ability to resolve field-induced changes in $\kappa /T$. The statistical
  error in the fit of $\kappa_0/T$ is much smaller than 3 $\mu
  $W~K$^{-2}$~cm$^{-1}$ and given as the error bars in Figs.~\ref{fig:kvsH} and
  \ref{fig:reswkvsH}b.}
\end{thebibliography}

\end{document}